\documentclass[prl,twocolumn,superscriptaddress]{revtex4}

\usepackage{graphicx}

\begin{document}
\newtheorem{theorem}{Theorem}
\newtheorem{acknowledgement}[theorem]{Acknowledgement}
\newtheorem{algorithm}[theorem]{Algorithm}
\newtheorem{axiom}[theorem]{Axiom}
\newtheorem{claim}[theorem]{Claim}
\newtheorem{conclusion}[theorem]{Conclusion}
\newtheorem{condition}[theorem]{Condition}
\newtheorem{conjecture}[theorem]{Conjecture}
\newtheorem{corollary}[theorem]{Corollary}
\newtheorem{criterion}[theorem]{Criterion}
\newtheorem{definition}[theorem]{Definition}
\newtheorem{example}[theorem]{Example}
\newtheorem{exercise}[theorem]{Exercise}
\newtheorem{lemma}[theorem]{Lemma}
\newtheorem{notation}[theorem]{Notation}
\newtheorem{problem}[theorem]{Problem}
\newtheorem{proposition}[theorem]{Proposition}
\newtheorem{remark}[theorem]{Remark}
\newtheorem{solution}[theorem]{Solution}
\newtheorem{summary}[theorem]{Summary}    
\def\r{{\bf{r}}}
\def\i{{\bf{i}}}
\def\j{{\bf{j}}}
\def\m{{\bf{m}}}
\def\k{{\bf{k}}}
\def\kt{{\tilde{\k}}}
\def\mt{{\hat{t}}}
\def\mG{{\hat{G}}}
\def\mg{{\hat{g}}}
\def\mGa{{\hat{\Gamma}}}
\def\mS{{\hat{\Sigma}}}
\def\mT{{\hat{T}}}
\def\K{{\bf{K}}}
\def\P{{\bf{P}}}
\def\q{{\bf{q}}}
\def\Q{{\bf{Q}}}
\def\p{{\bf{p}}}
\def\x{{\bf{x}}}
\def\X{{\bf{X}}}
\def\Y{{\bf{Y}}}
\def\F{{\bf{F}}}
\def\G{{\bf{G}}}
\def\bG{{\bar{G}}}
\def\mbG{{\hat{\bar{G}}}}
\def\M{{\bf{M}}}
\def\V{\cal V}
\def\tchi{\tilde{\chi}}
\def\tx{\tilde{\bf{x}}}
\def\tk{\tilde{\bf{k}}}
\def\tK{\tilde{\bf{K}}}
\def\tq{\tilde{\bf{q}}}
\def\tQ{\tilde{\bf{Q}}}
\def\si{\sigma}
\def\ep{\epsilon}
\def\hep{{\hat{\epsilon}}}
\def\al{\alpha}
\def\be{\beta}
\def\ep{\epsilon}
\def\bep{\bar{\epsilon}_\K}
\def\mep{\hat{\epsilon}}
\def\up{\uparrow}
\def\de{\delta}
\def\De{\Delta}
\def\up{\uparrow}
\def\dwn{\downarrow}
\def\ksi{\xi}
\def\etha{\eta}
\def\product{\prod}
\def\goto{\rightarrow}
\def\switch{\leftrightarrow}

\title{On the origin of the pseudogap in underdoped cuprates}
\author{
Th.A.~Maier, M.~Jarrell, A.~Macridin, and F.-C.~Zhang
}
\address{Department of Physics, University of Cincinnati, Cincinnati OH 45221, USA}

\begin{abstract}
We investigate the microscopic origin of the pseudogap in the weakly doped 
2D Hubbard model using Quantum Monte Carlo within the dynamical cluster 
approximation.  We compare our results with proposed scenarios for the 
pseudogap. All our numerical evidence is in favor of spin-charge separation 
as described in the resonating valence bond picture as the cause of the 
pseudogap behavior. Scenarios of  "preformed pairs", the coupling of 
quasiparticles to antiferromagnetic spin-fluctuations and stripes are 
inconsistent with our results.

\end{abstract}

\maketitle

\paragraph*{Introduction}
The existence of a pseudogap, i.e. a large suppression of low-frequency 
spectral weight in the normal state of underdoped high-temperature 
superconductors, is now a commonly accepted experimental fact. A multitude 
of experiments has probed the magnetic, thermodynamic, transport and optical 
properties of the underdoped cuprates (for an overview see 
Refs.~\cite{timusk,tallon}).

Some of the earliest indications of a pseudogap in the spin-channel were 
found in NMR-experiments \cite{warren}, where the spin-susceptibility as 
measured by the Knight-shift was seen to decrease with decreasing temperature
well above the superconducting  critical temperature $T_c$. Further 
indirect evidence of a pseudogap was found in specific heat 
measurements \cite{loram}. In underdoped samples, the electronic contribution 
starts to decrease with decreasing temperature in the normal state well above 
$T_c$. In transport measurements the crossover of the linear dependence of 
the ab-plane resistivity in temperature to a stronger dependence \cite{takagi} 
was taken as evidence for the opening of a pseudogap. However, the most direct 
and reliable measurements of the normal state pseudogap are angle-resolved 
photoemission experiments \cite{ding}. A highly anisotropic 
($d_{x^2-y^2}$-like) suppression of low-energy spectral weight persisting 
far above $T_c$ has been found in underdoped samples.  This pseudogap closes 
at a temperature $T^*$ consistent with the crossover temperatures determined 
from other measurements.

Theorists have responded with several scenarios\cite{kampf,emery,anderson,
varma,anderson2,zhang,rice} for the pseudogap.  Scenarios, based on the 
vicinity of the underdoped system to antiferromagnetic ordering, hold the 
coupling of quasiparticles to antiferromagnetic  spin fluctuations
responsible for the pseudogap behavior \cite{kampf}: As a consequence of 
short-ranged antiferromagnetic correlations, a shadow-band forms and a pseudogap opens 
up in the density of states.

Motivated by the fact that the pseudogap has the same symmetry as the
superconducting gap, "preformed pair" scenarios associate the pseudogap
with fluctuations that lead to $d$-wave superconductivity \cite{emery}.
In these scenarios the transition to the normal state is controlled by
the vanishing of the superfluid density and not the closing of the superconducting gap.
Due to strong pairing correlations above $T_c$, precursors of the superconducting gap
are seen as pseudogap in the low-energy excitations.

Ideas involving spin-charge separation are based on the resonating valence 
bond (RVB) picture \cite{anderson}. In these scenarios the pseudogap is due 
to d-wave singlet pairing of spin $1/2$, charge neutral fermions, called 
spinons.  At $T_c$ the holons, i.e. spin 0 charge excitations, become coherent 
and recombine with spinons to form electron pairs which renders the system superconducting.

Further scenarios are based on the existence of a quantum critical
point (QCP) close to optimal doping  that associate the pseudogap with a broken
symmetry (see eg. Ref.~\cite{varma}).  Other approaches that ascribe the
formation of the pseudogap to the presence of strong charge fluctuations
are motivated by the observation of charge stripes in some cuprate 
superconductors.

Presently, there is no consensus as to the origin of the pseudogap in the
underdoped cuprates so numerical investigations based on model calculations
are highly desirable. Early in the history of high-$T_c$ superconductors it was
realized that the two-dimensional (2D) Hubbard model in the intermediate coupling regime,
i.e. where the Coulomb interaction is of the order of the bandwidth, or closely related
models like the t-J-model, should capture the essential low-energy physics of the
cuprates \cite{anderson2}.

In this report we study the pseudogap behavior of the 2D Hubbard model using
the dynamical cluster approximation \cite{DCA_Hettler1,DCA_Hettler2,DCA_Maier1,DCA_Jarrell1} (DCA).
The DCA is a non-perturbative approach for the thermodynamic limit, which systematically
incorporates the effects of nonlocal correlations to local approximations like the dynamical
mean field approach (DMFA) \cite{DMFA}, by mapping the lattice onto a self-consistently
embedded cluster of size $N_c$. We solve the cluster problem using a combination of Quantum Monte Carlo (QMC) and the maximum entropy method to obtain dynamics \cite{DCA_Jarrell3}. Since further insight into the pseudogap phenomenon can be obtained by studying the effects of lattice frustration, we also present results
for the 2D Hubbard model with additional next nearest neighbor hopping $t^\prime$. We show that our results for the
pseudogap behavior in the 2D Hubbard model support the RVB picture and exclude other mechanisms for the pseudogap.

\paragraph*{Formalism}

A detailed discussion of the DCA formalism was given in previous
publications \cite{DCA_Hettler1,DCA_Hettler2,DCA_Maier1,DCA_Jarrell1, DCA_Jarrell3}.
The DCA is based on the assumption \cite{thurston} that the lattice
self-energy is only weakly momentum dependent and can be approximated by a constant
within each of a set of cells centered at a corresponding set of cluster $\K$-points.
The Green-functions used to calculate the self-energy are coarse-grained or averaged
over these cells. This greatly reduces the complexity of the lattice problem to that of a
periodic cluster embedded in a host which has to be determined self-consistently.
The DCA reduces to the DMFA for a cluster size of one and becomes exact when the cluster size is equal to the system size.

\paragraph*{Results}

We present results of DCA calculations for the conventional 2D Hubbard model
characterized by an overlap integral  $t$ between nearest neighbors and $t^\prime$ between next nearest neighbors  and an on-site Coulomb repulsion $U$.
We set $t=0.25$ and choose the magnitude $\left|t^\prime/t\right|\leq 0.5$,  so that the band-width $W=8t=2$, and study the intermediate
coupling regime $U=W$.  We study the initial corrections to the DMFA by setting
the cluster size to $N_c=4$, the smallest cluster size  which allows
for d-wave pairing. We have previously shown that this cluster size is large 
enough to capture the qualitative low-energy physics of the cuprate 
superconductors \cite{DCA_Jarrell2, DCA_Maier2}. Due to numerical restrictions 
we are currently not able to perform systematic studies with increasing cluster 
size. The $N_c=4$ simulations presented in this paper should therefore be 
interpreted as qualitative extended mean-field results that describe effects 
on short-length scales (within the cluster).
\begin{figure}[h]
\includegraphics*[width=3.in]{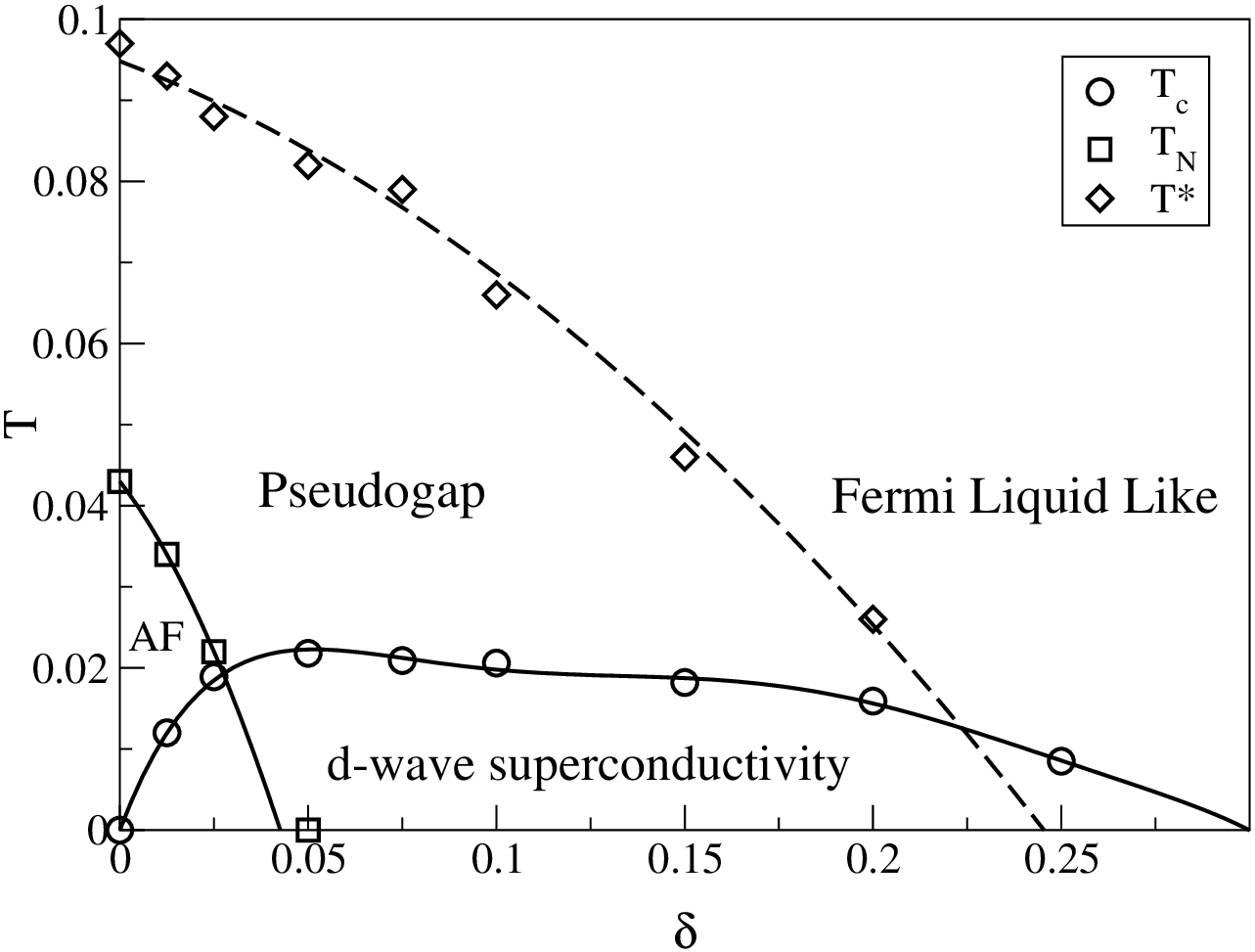}
\caption{Temperature-doping phase diagram of the 2D Hubbard model when $U=W=2$, $t^\prime=0$ and $N_c=4$.  The critical temperature $T_c$ and Ne\'{e}l-temperature 
$T_{\rm N}$ were obtained from the divergence of the pair-field- and staggered 
spin-susceptibilities, the crossover temperature $T^*$ from the maximum in the 
uniform spin-susceptibility.}
\label{fig:pd}
\end{figure}

\begin{figure}[h]
\includegraphics*[width=3.in]{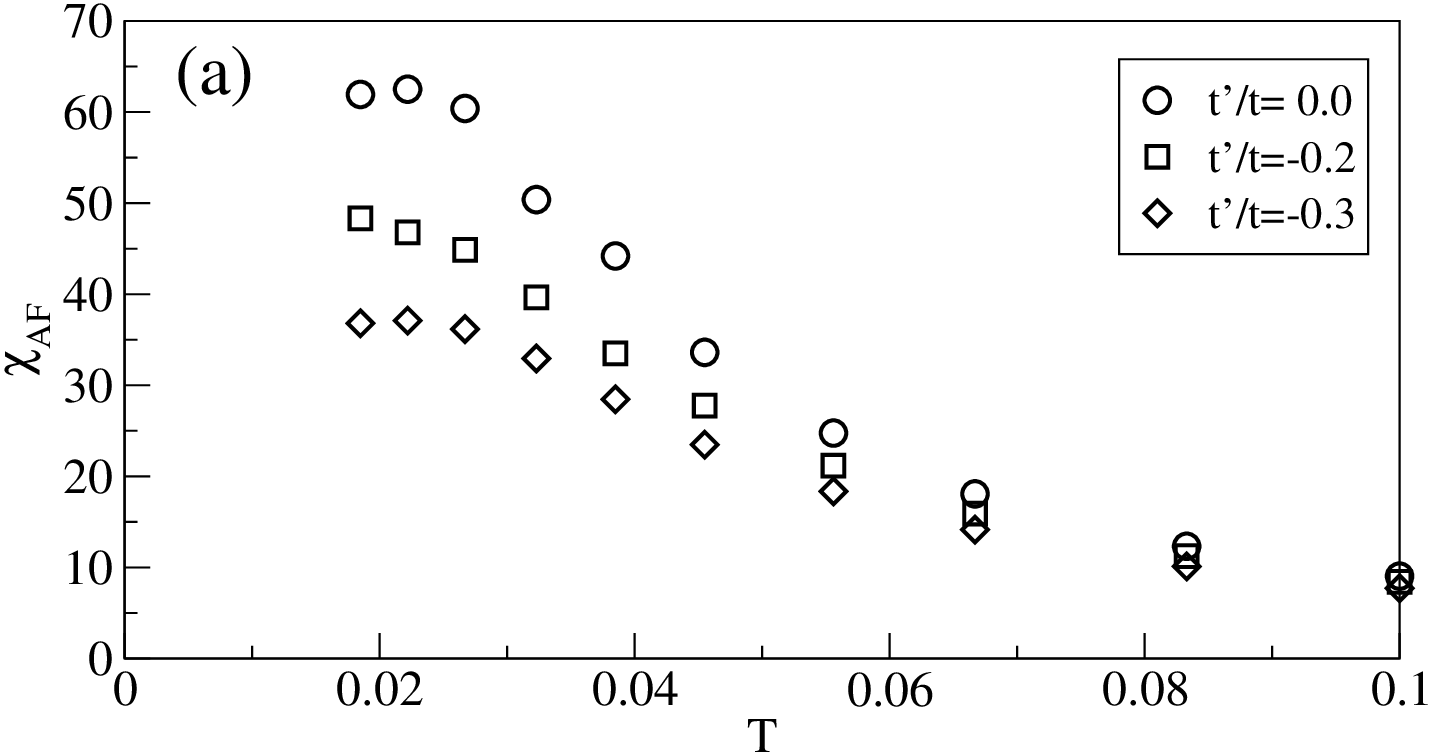}
\includegraphics*[width=3.in]{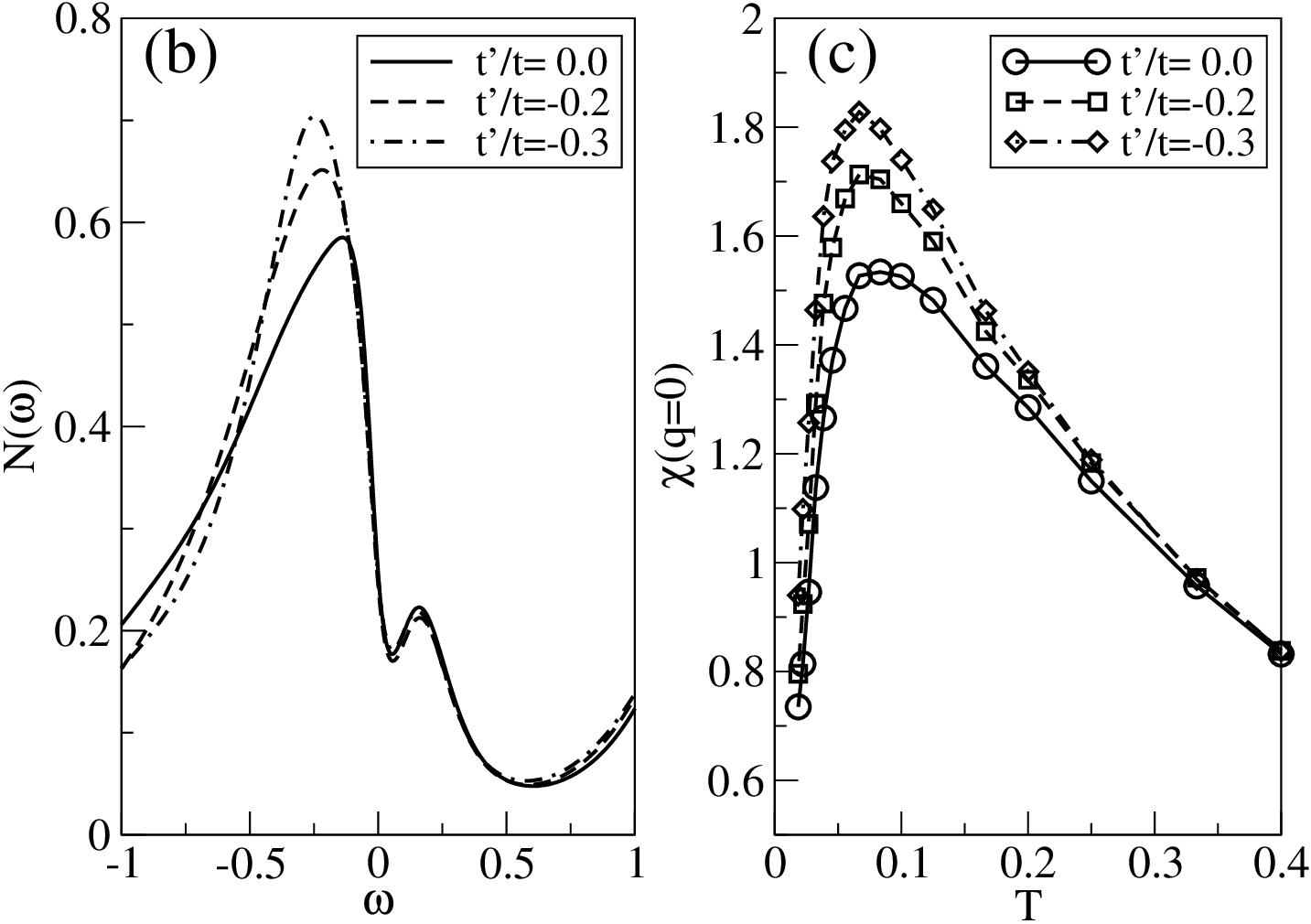}
\caption{(a) The antiferromagnetic spin-susceptibility versus temperature, (b) the density of states at fixed temperature $T=0.022$ in the pseudogap regime and (c) the uniform magnetic susceptibility for $\delta=0.05$ and  for different values of the next-nearest neighbor hopping amplitude $t^\prime$. Frustration suppresses antiferromagnetic correlations but does not affect the pseudogap.}
\label{fig:chiQ}
\end{figure}

Our results are summarized in the temperature-doping ($T$-$\delta$) phase diagram shown
in Fig.\ref{fig:pd}. We find regions of antiferromagnetism, d-wave superconductivity,
pseudogap and Fermi-liquid like behaviors. In this report we focus on the pseudogap
behavior found at low doping $\delta$. The pseudogap region is characterized by a suppression
of spin excitations \cite{moreo} below a crossover temperature $T^*$, defined via the maximum
in the spin-susceptibility (see Fig.\ref{fig:chiQ}c for $t^\prime=0.0$ and
doping $\delta=0.05$) when accompanied by the formation of a pseudogap
in the density of states for temperatures $T<T^*$ (see Fig.\ref{fig:chiQ}b). $T^*$ at
zero doping is of the order of the magnetic exchange coupling $J=4t^2/U=0.125$.
To a very good approximation, $T^*$ is equal to the mean-field Ne\'el temperature.
These observations indicate that {\em{short-ranged antiferromagnetic correlations are a prerequisite to the pseudogap}}.

Thus, it is natural to explore the scenario in which the antiferromagnetic short-range order
causes the pseudogap. To this end we study the effects of lattice frustration, i.e. of the magnitude of the next nearest neighbor hopping integral $t^\prime$ on the pseudogap. With increasing $\left| t^\prime\right|$ the spins on the  lattice become frustrated and antiferromagnetic correlations are expected to be suppressed. On the other hand,  we expect a possible RVB spin-liquid state to be essentially resistant to, or even be stabilized by finite values of $t^\prime$. In Ref.\cite{luca} it was shown that the ground  state of the frustrated antiferromagnetic Heisenberg model is almost exactly reproduced by a RVB wave function.  In  Fig.~\ref{fig:chiQ}a we plot our results for the antiferromagnetic susceptibility for different values of $t^\prime$. Clearly antiferromagnetic correlations are suppressed as a function of increasing magnitude of $t^\prime$. 

On the other hand, the density of states plotted in Fig.~\ref{fig:chiQ}b  remains  unchanged near the Fermi surface with increasing magnitude of $t^\prime$. In addition, the temperature  $T^*$ where the anomaly in the uniform magnetic susceptibility shown in Fig.~\ref{fig:chiQ}c occurs, is essentially unaffected by the value of $t^\prime$. Lattice frustration has little effect on the pseudogap.
Thus we conclude that short-ranged antiferromagnetic correlations 
{\em{alone}} cannot be responsible for the evolution of the pseudogap, 
inconsistent with scenarios based on the coupling of quasiparticles to 
antiferromagnetic spin-fluctuations \cite{kampf}. Moreover, we infer that these results favor RVB physics over antiferromagnetic short-range order as the origin of the pseudogap.   

\begin{figure}[h]
\includegraphics*[width=3.5in]{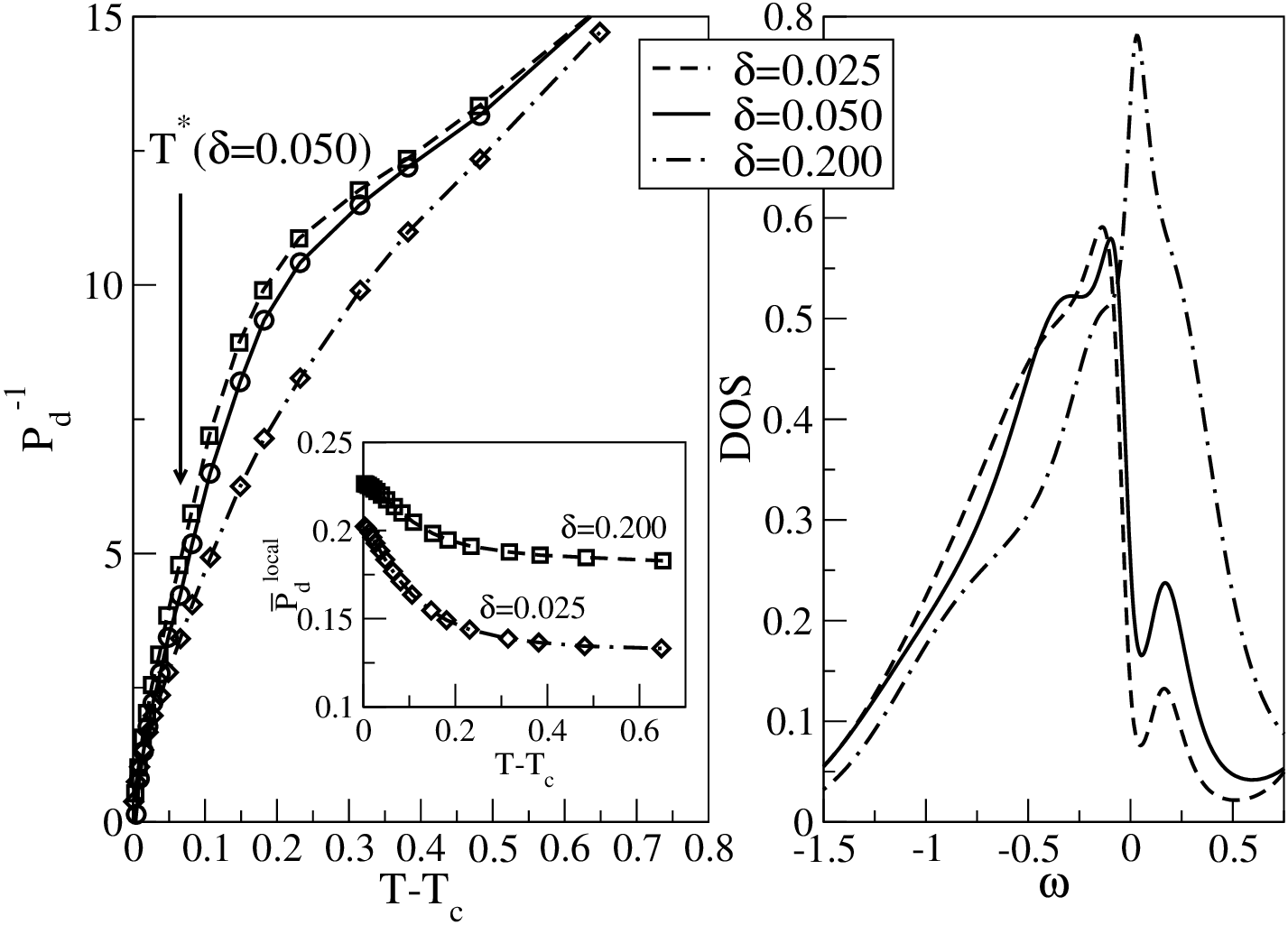}
\caption{Left: The inverse pairfield-susceptibility (inset: local cluster 
equal-time d-wave pairfield-susceptibility) versus reduced temperature for 
various dopings for $t^\prime=0$. Right: The density of states for the same doping levels at 
$T=0.022$.}
\label{fig:pc}
\end{figure}

In the "preformed pairs" scenarios the pseudogap is ascribed to fluctuations 
that lead to superconductivity at a lower temperature $T_c$. In this case we 
would expect a corresponding signature in the d-wave pair-field susceptibility, 
i.e. an enhancement due to pairing correlations in the pseudogap region.  In 
the left panel of Fig.~\ref{fig:pc} we compare the inverse d-wave pair-field 
susceptibility $P_d^{-1}$ of the clean system ($x=0$) as a function of reduced 
temperature $T-T_c$ for different dopings  $\delta$ in the pseudogap 
($\delta=0.025, 0.05$) and overdoped ($\delta=0.20$) regions.  Here we are 
interested in temperatures below $T^*$ indicated by the arrow for 
$\delta=0.05$.  As the doping increases from the pseudogap to the overdoped 
region, $P_d^{-1}$ decreases at fixed reduced temperature indicating that 
pairing correlations in the overdoped region are even more pronounced than 
in the pseudogap region. Our result in the inset of Fig.\ref{fig:pc} for the
local cluster \cite{cluster} equal-time d-wave pair-field susceptibility
$\bar{P}_d^{local}=\frac{1}{4N_c}\sum_{ijk} g_{ij}g_{ik}
\langle c^{}_{i\uparrow}c^{}_{j\downarrow}c^\dagger_{k\downarrow}
c^\dagger_{i\uparrow}\rangle$, where $g_{ij}$ is the Fourier-transform of 
$\cos k_x-\cos k_y$, supports this finding. This quantity does not contain 
phase fluctuations and thus can be taken as direct evidence against strong 
pairing correlations at underdoping.    These results lead us to eliminate 
the "preformed-pairs" scenarios as an origin for the pseudogap in our results.

Among other scenarios for the pseudogap include the idea of QCP, stripes, and 
spin-charge separation.  In most scenarios, the QCP exists at the zero 
temperature terminus of $T^*$ versus doping.  However, near this point, we 
find no evidence for a QCP in our simulations; i.e. none of the many
susceptibilities that we measure are enhanced as we approach this point.
We do not see any signatures of stripes in the charge susceptibility (not 
shown). However, stripes might be important for larger clusters.

As already discussed, the results for the frustrated system are  consistent with a 
RVB, i.e.\ spin-charge separated state. In order to address the question of 
spin-charge separation unambiguously, we would have to calculate the lattice 
dynamic spin- and charge- structure factors for a range of momenta and extract 
the spin and charge dispersions. However, these calculations, although formally 
possible, are presently numerically too expensive to perform.  Nevertheless, 
we believe that we can gain insight into this question by studying the behavior 
of the corresponding quantities calculated on the cluster only\cite{cluster}.
Preliminary results (not shown here) for the dynamic cluster spin- and charge 
susceptibilities at low doping are consistent with spin-charge separation: The 
spin-susceptibility, dominated by fluctuations at $\Q=(\pi,\pi)$, becomes 
suppressed at low frequencies with decreasing temperature well above $T_c$. In 
the charge-susceptibility, which is strongest at $\q=(0,0)$, weight builds  
at low frequencies as the temperature is lowered and no sign of a pseudogap 
is seen.

\begin{figure}[h]
\includegraphics*[width=3.2in]{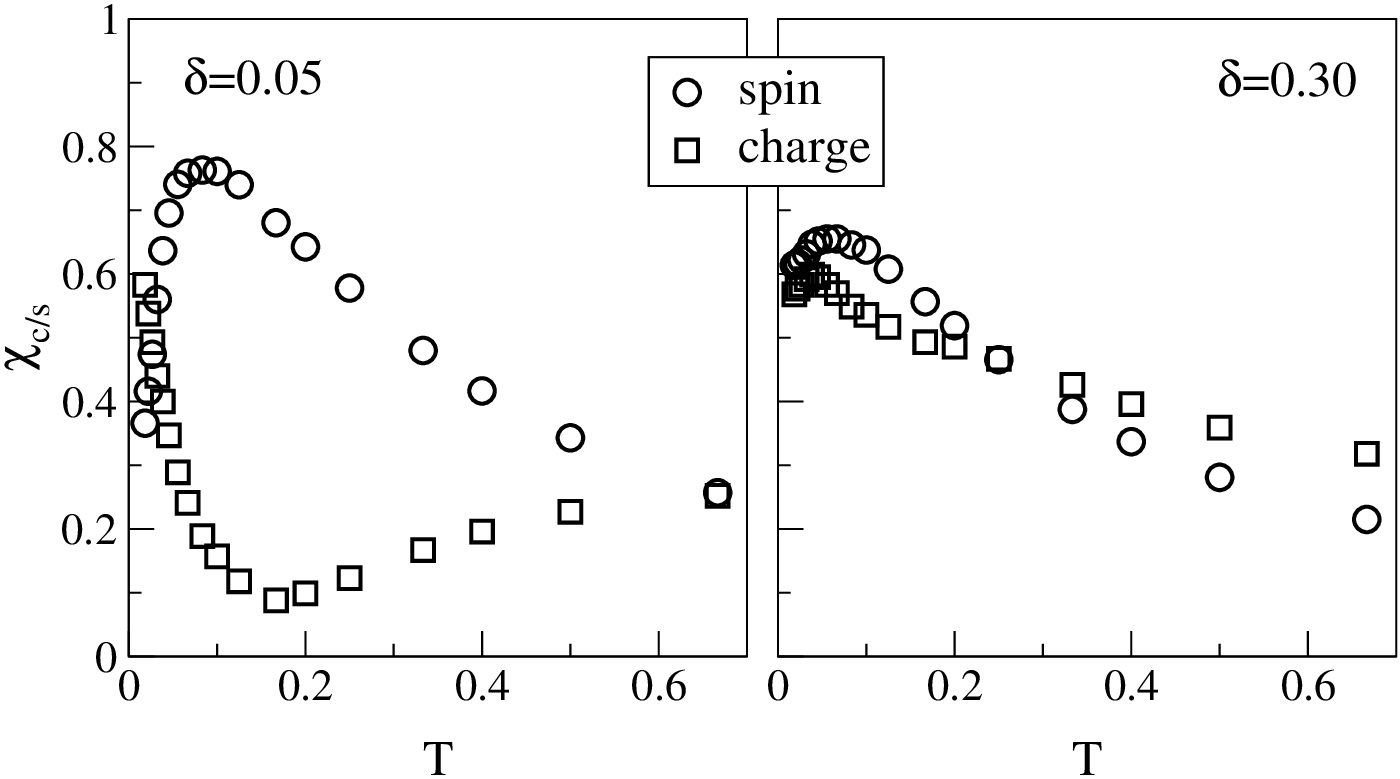}
\caption{Temperature dependence of the uniform spin and charge susceptibility
in the underdoped (left) and overdoped (right) regime when $U=W=2$, $t^\prime=0$ and $N_c=4$.}
\label{fig:chisc}
\end{figure}

The results shown in Fig.~\ref{fig:chisc} for the static uniform lattice 
spin- (circles) and charge-susceptibilities (squares)  versus temperature at 
underdoping ($\delta=0.05$, left) and overdoping ($\delta=0.30$, right) support 
this picture. As already discussed, the spin-susceptibility $\chi_s$ in the 
underdoped system shows an anomaly at a temperature $T^*$ below which 
spin-excitations are suppressed. The charge-susceptibility $\chi_c$ in the 
underdoped system however displays qualitatively different behavior: When the 
spin degrees of freedom become suppressed at $T\lesssim T^*$, $\chi_c$ starts 
to rise with decreasing temperature.  This clearly  indicates that coherence 
in the charge channel starts to set in with decreasing temperature  whereas 
the spin excitations become suppressed.  These results are consistent with 
recent angle-resolved photoemission experiments \cite{Ding2} where superconducting order 
in the underdoped system was found to be accompanied by an emerging
quasiparticle coherence as seen in the spectra.

Thus, our results for the underdoped region can be interpreted within an RVB,
spin-charge separated picture, where the development of the pseudogap originates in
the  pairing of the spin degrees of freedom and superconductivity is driven by
coherence of the charge excitations. In the overdoped region (right panel in Fig.~\ref{fig:chisc})
our results show that the system becomes more conventional, i.e. spin and charge excitations behave in a similar way.

\paragraph*{Summary}
We have investigated the microscopic origin of the pseudogap in DCA simulations of the
underdoped 2D Hubbard model.  We find that short-ranged correlations alone, although necessary
for the pseudogap to emerge, is not sufficient to describe its origins.  Similarly, although stripe fluctuations may emerge for larger clusters, no indications were found in our calculations, and therefore stripes are not the origin of the pseudogap.
Likewise, we find no indication for strong pairing fluctuations for temperatures $T_c<T<T^*$,
in disagreement with pre-formed pairing scenarios.  Finally, although we cannot exclude a quantum
critical point at finite doping as an origin, we find no evidence for it.
Instead, all of our numerical evidence points to spin-charge separation as
described by Anderson's RVB theory as the microscopic origin of the pseudogap.
In particular, in the pseudogap region, spin degrees of freedom are suppressed
while charge excitations become more coherent with decreasing temperature.
At overdoping where no pseudogap is found, spin and charge behave qualitatively similar.

\paragraph*{Acknowledgments}  This work was supported by NSF grant DMR-0073308 and by NSF cooperative
agreement ACI-9619020 through computing resources provided by the NPACI at the Pittsburgh Supercomputer Center.



\begin{references}
\bibitem{cluster} Correlation functions calculated on the cluster are not identical to their lattice counterparts. Only irreducible quantities correspond one to one. Nevertheless we expect the cluster quantities to be qualitatively similar to the corresponding lattice properties.
\end{references}
\end{document}